\documentclass{article}

\usepackage{amsmath}
\usepackage{amssymb}
\usepackage[top=1in, bottom=1in, left=1.5in, right=1.5in]{geometry}
\usepackage{graphics}
\usepackage{hyperref}
\usepackage[pdftex]{color}
\usepackage[pdftex]{graphicx}
\usepackage{textpos}

\begin{document}

	\title{Quasilocal Energy in FRW Cosmology}
	\author{M. M. Afshar\\
		\textit {\small{Department of Physics, University of California}}\\
		\textit {\small{Davis, CA 95616, USA}}\\
		\textit {\small{email:  mafshar@physics.ucdavis.edu}}}
	\date{August 3, 2009}
	\maketitle
	
	\begin{abstract}
		This paper presents a calculation of the quasilocal energy of a generic FRW model of the universe.  The results have the correct behavior in the small-sphere limit and vanish for the empty Milne universe.  Higher order corrections are found when comparing these results to classical calculations of cosmological energy.  This case is different from others in the literature chiefly in that it involves a non-stationary spacetime.  This fact can be used to differentiate between the various formulations of quasilocal energy.  In particular, the formulation due to Brown and York is compared to that of Epp.  Only one of these is seen to have the correct classical limit.
	\end{abstract}

	\section{Introduction}
	\label{Introduction}
	In general relativity, as other areas of physics, the notion of energy is of both conceptual and practical importance.  Unlike other areas of physics however, a sensible and universal definition of energy has been difficult to come by.  Difficulties in defining energy in the context of general relativity became apparent from the very beginning.  Einstein's pseudo-tensor formulation, as well as the many variations that have appeared since \cite{Landau:1975, Szabados:2004vb}, are not entirely satisfactory because they are not generally covariant.  The global definitions of energy, such as the Bondi or ADM formulations, also seem inadequate because they are not localized and must necessarily refer to the entire spacetime.\footnote{See however \cite{Bartnik:1989zz} for a mass function that `localizes' the ADM energy, albeit in a computationally intractable manner.}  A reasonable compromise seems to have been reached with the notion of `quasilocal energy' -- energy that is defined for an extended but finite region of spacetime.  There is a plethora of definitions of quasilocal energy, many of which are thoroughly reviewed in \cite{Szabados:2004vb}.
	
	The virtues and weaknesses of these definitions can be revealed by application to specific spacetimes.  The Schwarzschild and Kerr spacetimes are common testing grounds for the various notions of quasilocal energy.  However the fact that these spacetimes are stationary limits their discriminating power.  Non-stationary examples can discern among the various notions of quasilocal energy in a way that stationary examples cannot.  The primary purpose of this paper is to examine the case of a generic Friedmann-Lema\^itre-Robertson-Walker universe, commonly referred to as FRW cosmology in much of the literature.  Aside from its practical applicability, this non-stationary example reveals in a new way how various notions of quasilocal energy differ from one another.  Due to limitations of time and space, I have chosen to focus on only two formulations of quasilocal energy among the many possible ones.  The first is the formulation of Brown and York \cite{Brown:1992br} and the second is the closely related formulation of Epp \cite{Epp:2000zr}.  I will comment on other formulations only in passing, although they certainly deserve greater attention.
	
	From a theoretical point of view, the Brown-York formulation is appealing because it is based on the rigorous Hamilton-Jacobi theory of classical mechanics.  Although one may question the applicability of this theory to general relativity, the ends ultimately justify its use.  So this formulation will be my starting point in Sec. \ref{Quasilocal Energy a la Brown and York}.  Calculation of the Brown-York quasilocal energy, $E_{BY}$, for the the FRW model is carried out in Sec. \ref{FRW Cosmology a la Brown and York}, where it is seen to be contradictory to physical expectations.  In Sec. \ref{Quasilocal Energy a la Epp} the Epp quasilocal energy, $E_{E}$, is introduced, and the calculation for the FRW model is repeated in Sec. \ref{FRW Cosmology a la Epp}.  It will be shown that in the appropriate limits, this second result provides a reasonable and expected value of energy for an FRW universe.

	\section{Quasilocal Energy \`a la Brown and York}
	\label{Quasilocal Energy a la Brown and York}
\begin{figure}[htbp]
		\begin{center}
			\includegraphics[width=2in]{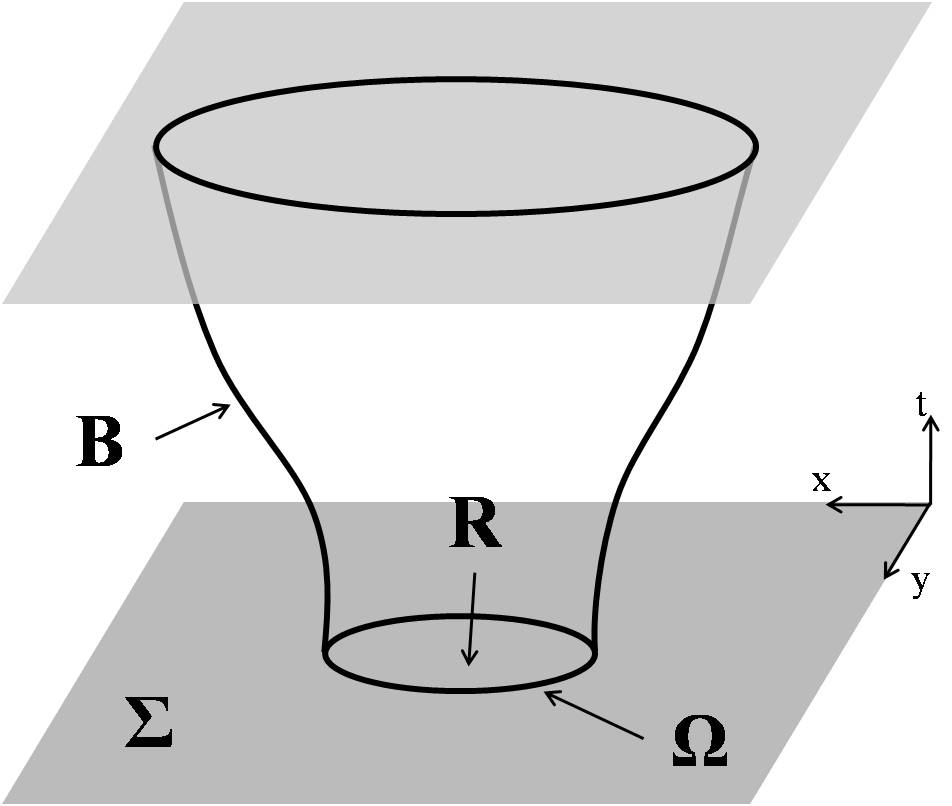}
			\caption{Spacetime Diagram}
			\label{Pic1.jpg}
		\end{center}
	\end{figure}

	To begin with, the notation will be as follows.  $M$ denotes the spacetime manifold.  I assume that $M$ has topology $\mathbb{R} \times \Sigma$, where $\Sigma$ is a fixed three-dimensional manifold.  This assumption allows the foliation of $M$ into spacelike leaves, $\Sigma_t$, each of which represents space at one instant in time.  To be more precise, I assume that there exists a diffeomorphism $f:\mathbb{R} \times \Sigma \rightarrow M$.  Then for $t \in \mathbb{R}$ and $\sigma \in \Sigma$, $f \left( t,\sigma \right)$ represents one point of the leaf $\Sigma_t$.  Note that $f = \left( f^0,f^1,f^2,f^3 \right)$, and that it can be used to define a vector field $t^\mu \equiv \partial f^\mu / \partial t$ on $M$.  This vector field specifies the direction of time evolution at each point.  $R$ denotes a region in $\Sigma_t$ for some fixed $t$, and $\Omega$ denotes the boundary of this region, $\Omega = \partial R$.  $B$ denotes the manifold that results when $\Omega$ evolves in time according to the vector field $t^\mu$.  In other words, $B$ denotes the manifold that results by moving points of $\Omega$ along the integral curves of $t^\mu$.  Note that $B$ will have the topology $\mathbb{R} \times \Omega$.\footnote{The notation here closely resembles that of Brown and York \cite{Brown:1992br}, with the exception that where I use $B$ they use $^3B$, and where I use $\Omega$ they use $B$.  Epp \cite{Epp:2000zr} uses similar notation, with the exception that where I use $\Omega$ he uses $S$.}  See Figure \ref{Pic1.jpg} for a simplified depiction of the geometry.
	
	As in \cite{Brown:1992br}, each of the manifolds described above is endowed with a metric:  $g_{\mu\nu}$ is the metric on $M$, while $h_{ij}$, $\gamma_{ij}$ and $\sigma_{ab}$ are the induced metrics on $\Sigma$, $B$ and $\Omega$ respectively.  Note the use of indices, where $\mu, \nu = 0,1,2,3$ are used to index coordinates on $M$, $i,j = 1,2,3$ are used (somewhat confusingly) to index coordinates on either $\Sigma$ or $B$, and $a,b = 1,2$ are used to index coordinates on $\Omega$.  The unit normal to $\Sigma$ in $M$ is denoted by $u_\mu$, and the extrinsic curvature with respect to this normal is denoted by $K_{\mu \nu}$.  The unit normal to $B$ in $M$ is denoted by $n_\mu$, and the extrinsic curvature with respect to this normal is denoted by $\Theta_{\mu \nu}$.  The unit normal to $\Omega$ in $\Sigma$ is denoted by $n_i$, and the extrinsic curvature with respect to this normal is denoted by $k_{ij}$.  Table \ref{table - summary of notation} summarizes some of this notation.
	\begin{table}[htbp]
		\begin{displaymath}
			\begin{array}{|c|c|c|c|c|}
				\hline
				Manifold & Dims. & Metric      & Normal  & Ext. Curv.       \\
				\hline \hline
				M			   & 4     & g_{\mu \nu} &         &                  \\
				\Sigma	 & 3     & h_{ij}      & u_{\mu} & K_{\mu \nu}      \\
				B			   & 3     & \gamma_{ij} & n_{\mu} & \Theta_{\mu \nu} \\
				\Omega	 & 2     & \sigma_{ab} & n_{i}   & k_{ij}           \\
				\hline
			\end{array}
		\end{displaymath}
		\caption{Summary of Notation}
		\label{table - summary of notation}
	\end{table}
	
	The main result of \cite{Brown:1992br} is that a generalized energy-momentum tensor, $\tau^{ij}$, may be defined which describes the \emph{total} energy, i.e. energy of matter content as well as the gravitational energy, inside an arbitrary region.  The motivation for this definition comes from the Hamilton-Jacobi formalism in classical mechanics.  If $S_{clas}$ denotes the action calculated for a classical solution, i.e. a solution satisfying the equations of motion, then momentum can be expressed as the derivative of $S_{clas}$ with respect to coordinates on the \emph{boundary}.  Similarly, Brown and York define the generalized energy-momentum tensor $\tau^{ij}$ as:
	\begin{align}
		\label{generalized energy-momentum definition}
		\tau^{ij} \equiv \frac{2}{\sqrt{|\gamma|}} \frac{\delta S_{clas}}{\delta \gamma_{ij} \left|_{_{B}} \right.} \,\, .
	\end{align}
	Here $S_{clas}$ is the total action for gravity and matter calculated for a classical solution, and the variation is with respect to the metric on the boundary $B$.  Calculating the functional derivative in \eqref{generalized energy-momentum definition} shows that \cite{Brown:1992br}:
	\begin{align}
		\tau^{ij} &= -\frac{2}{\sqrt{|\gamma|}} \pi^{ij},
	\end{align}
	where $\pi^{ij} \equiv \delta S / \delta \dot{\gamma}_{ij}$ is the canonical momentum conjugate to the boundary metric.  The generalized energy-momentum tensor is used to define the surface energy density $\epsilon$, surface momentum density $j^a$, and surface stress density $s^{ab}$ as follows:
	\begin{align}
		\epsilon &\equiv u_i u_j \tau^{ij} \\
		j^a &\equiv -\sigma^a_i u_j \tau^{ij} \\
		s^{ab} &\equiv \sigma^a_i \sigma^b_j \tau^{ij}.
	\end{align}
	With the appropriate choice of coordinates, the generalized energy-momentum tensor can be written as:
	\begin{align}
		\tau^{ij} &= \left[
			\begin{array}{c|c}
				\epsilon & j^a \\
				\hline
				j^a & s^{ab}
			\end{array}
			\right].
	\end{align}
	The quasilocal energy inside the region bounded by $\Omega$ is then defined as:
	\begin{equation}
		\label{BY QLE Definition}
		E \equiv \int_\Omega d^2x \sqrt{|\sigma|} \,\, \epsilon.
	\end{equation}
	
	To simplify equation \eqref{BY QLE Definition} note first that, using the same mathematics as in the ADM formalism, one can express the canonical momentum as:
	\begin{equation}
		\pi^{ij} = -\frac{1}{16 \pi G} \sqrt{|\gamma|} \left( \Theta \gamma^{ij} - \Theta^{ij} \right),
	\end{equation}
	where $\Theta_{ij}$ is the extrinsic curvature of $B$ as a submanifold of $M$.  The surface energy density can then be written as:
	\begin{align}
		\epsilon \equiv u_i u_j \tau^{ij} = -\frac{2}{\sqrt{|\gamma|}} u_i u_j \pi^{ij} = \frac{1}{8 \pi G} u_i u_j \left( \Theta \gamma^{ij} - \Theta^{ij} \right).
	\end{align}
	Decomposing $\Theta_{ij}$ into components tangent and normal to $\Sigma$, it can be shown that $\epsilon = k/8 \pi G$, where $k$ is the trace of the extrinsic curvature of $\Omega$ as a submanifold of $\Sigma$.\footnote{It should be emphasized that $k = \sigma^{ab} k_{ab}$ where $k_{ab}$ is the extrinsic curvature of $\Omega$ viewed as a submanifold of $\Sigma$.  One can define another extrinsic curvature for $\Omega$ viewed as a submanifold of $B$.  This point will prove to be important.}  This in turn allows one to write equation \eqref{BY QLE Definition} as:
	\begin{equation}
		\label{BY QLE Definition 2}
		E = \frac{1}{8 \pi G} \int_\Omega d^2x \sqrt{|\sigma|} \,\, k.
	\end{equation}
	
	As with most definitions of energy, a reference point must be chosen.  Since it is generally believed that the energy in Minkowski spacetime must be zero, the standard approach is to use this spacetime as the reference.  Equation \eqref{BY QLE Definition 2} is thus modified to yield the Brown-York quasilocal energy:
	\begin{align}
		\label{BY QLE Definition 3}
		E_{BY} &= E_{phys} - E_{ref} = \frac{1}{8 \pi G} \int_\Omega d^2x \sqrt{|\sigma|} \,\, k - \frac{1}{8 \pi G} \int_{\bar{\Omega}} d^2\bar{x} \sqrt{|\bar{\sigma}|} \,\, \bar{k},
	\end{align}
	where quantities with an overbar refer to the reference spacetime and those without refer to the physical spacetime.  Specifically, $\bar{\Omega}$ is a suitably chosen 2-manifold in Minkowski spacetime, and $\bar{k}$ is the trace of the extrinsic curvature of that 2-manifold.  There remains some ambiguity in the choice of $\bar{\Omega}$.  In \cite{Brown:1992br} it is suggested that $\bar{\Omega}$ ought to be an isometric embedding of $\Omega$ into Minkowski spacetime.  This prescription can be applied to a variety of simple examples with very good results.  For example, it can be applied to static, spherically symmetric spacetimes \cite{Brown:1992br}, and the results are in agreement with expectations.  However, when applied to the \emph{non-stationary} case of a Robertson-Walker spacetime, the results do not seem to be correct as demonstrated in the following subsection.

	\subsection{FRW Cosmology \`a la Brown and York}
	\label{FRW Cosmology a la Brown and York}
	FRW cosmology is described in comoving coordinates $\left( t,r,\theta,\phi \right)$ by the Robertson-Walker metric:
	\begin{equation}
		\label{FRW metric}
		ds^2 = -dt^2 + a^2(t) \left[ \frac{dr^2}{1-\kappa r^2} + r^2 d\theta^2 + r^2 \sin^2\theta \, d\phi ^2 \right],
	\end{equation}
	where $\kappa \in \{-1,0,+1\}$ is proportional to the spatial curvature and $a(t)$ is the dimensionless scale factor.
	
	To simplify the calculation, region $R$ is chosen to be spherically symmetric.  The boundary of this region, $\Omega$, is then a sphere of comoving radius $r$ at time $t$.  Adopting the natural constant-$t$ foliation of $M$, one may calculate the extrinsic curvature of $\Omega$ in $\Sigma$:
	\begin{align}
		k_{ij} &= \textnormal{diag} \left[  0 \,\, , \,\, -r \sqrt{1-\kappa r^2} \, a(t) \,\, , \,\, -r \sqrt{1-\kappa r^2} \, \sin^2\theta \, a(t)  \right].
	\end{align}
	The first term in \eqref{BY QLE Definition 3} is then calculated to be
	\begin{align}
		\label{BY QLE FRW physical term}
		E_{phys} &= \frac{1}{8 \pi G} \int_\Omega d\theta d\phi \,\, r^2 \sin\theta \, a^2(t) \left( \frac{-2 \sqrt{ 1-\kappa r^2}}{r \,\, a(t)} \right) = -\frac{1}{G} r \sqrt{1-\kappa r^2} \, a(t).
	\end{align}
	
	Alternatively, one can take a more direct approach and calculate the generalized energy-momentum tensor:
	\begin{align}
		\label{generalized energy-momentum tensor FRW}
		\tau_{ij} &= -\frac{1}{8 \pi G} \left( \Theta \gamma^{ij} - \Theta^{ij} \right) \nonumber \\
		          &= \textnormal{diag} \left[  -\frac{\sqrt{1-\kappa r^2}}{4 \pi G r a(t)} \,\, , \,\, \frac{r \sqrt{1-\kappa r^2} \, a(t)}{8 \pi G} \,\, , \,\, \frac{r \sqrt{1-\kappa r^2} \, \sin^2\theta a(t)}{8 \pi G}  \right].
	\end{align}
	The t-t component of this tensor is the surface energy density $\epsilon$ which, when integrated over $\Omega$, provides the same result as \eqref{BY QLE FRW physical term}:
		\begin{align}
		\label{BY QLE FRW physical term alternative}
		E_{phys} &= \int_\Omega d^2x \sqrt{|\sigma|} \,\, \epsilon = -\frac{1}{G} r \sqrt{1-\kappa r^2} \, a(t).
	\end{align}
	
	To calculate the reference term, I choose Minkowski spacetime as the reference spacetime and denote all quantities in the reference spacetime with a bar over them.  So Minkowski spacetime, $\bar{M}$, is coordinatized in spherical coordinates by $\bar{x}^\mu = \left( \bar{t},\bar{r},\bar{\theta},\bar{\phi} \right)$, and has the metric
	\begin{align}
		\bar{g}_{\mu \nu} &= \textrm{diag} \left[ -1 \,\, , \,\, 1 \,\, , \,\, \bar{r}^2 \,\, , \,\, \bar{r}^2 \sin^2\bar{\theta} \right].
	\end{align}
	Before calculating the reference term, one must find an isometric embedding, $\psi:  \Omega \rightarrow \bar{M}$, of the boundary $\Omega$ in Minkowski spacetime.  I use the following ansatz which preserves spherical symmetry:
	\begin{align}
		\label{isometric embedding ansatz}
		\psi \left( \theta,\phi \right) = \left( \bar{t}(t,r) \,\, , \,\, \bar{r}(t,r) \,\, , \,\, \theta \,\, , \,\, \phi \right).
	\end{align}
	The functions $\bar{t}(t,r)$ and $\bar{r}(t,r)$ are to be determined by the requirement that the metric be preserved, i.e. that the embedding be isometric.  This embedding defines $\bar{\Omega}$.  The metric on $\bar{\Omega}$ is the induced metric, i.e. the pull-back of Minkowski metric to the submanifold $\bar{\Omega}$:
	\begin{align}
		\bar{\sigma} \equiv \psi^* (\bar{g}) \quad \Rightarrow \quad \bar{\sigma}_{ab} = \textrm{diag} 
		\left[ \bar{r}^2 \,\, , \,\, \bar{r}^2 \sin^2\theta \right]
	\end{align}
	To enforce the requirement of isometry, the metrics on $\Omega$ and $\bar{\Omega}$ are equated, $\sigma_{ab} = \bar{\sigma}_{ab}$, with the result that $\bar{r} \left( t,r \right) = a(t) r$.  Note that equating the metrics does not determine $\bar{t}(t,r)$, although this ambiguity will not affect any of the following results.\footnote{If one were interested, there is a way to determine $\bar{t}(t,r)$ as well.  Instead of embedding $\Omega$ in $\bar{M}$ to obtain $\bar{\Omega}$, one could choose to embed $B$ in $\bar{M}$, and then take the appropriate cross-section to obtain $\bar{\Omega}$.  This procedure would provide an explicit expression for $\bar{t}(t,r)$.  Calculation of the quasilocal energy however would be the same in either case.}  Now the second term in \eqref{BY QLE Definition 3} can be calculated:
	\begin{align}
		\label{BY QLE FRW reference term}
		E_{ref} &= \frac{1}{8 \pi G} \int_{\bar{\Omega}} d\bar{\theta} d\bar{\phi} \,\, \bar{r}^2 \sin \bar{\theta} \left( \frac{-2}{\bar{r}} \right) = -\frac{1}{G} \bar{r} = -\frac{1}{G} r a(t).
	\end{align}
	
	Subtracting \eqref{BY QLE FRW reference term} from \eqref{BY QLE FRW physical term}, the Brown-York quasilocal energy of a Robertson-Walker spacetime is found to be:
	\begin{equation}
		\label{BY QLE FRW}
		E_{BY} = \frac{1}{G} r \left( 1 - \sqrt{1-\kappa r^2} \right) \, a(t).
	\end{equation}
	The obvious deficiency in this result is that for the case of a flat FRW model, $\kappa = 0$, the quasilocal energy is zero regardless of the matter content of the model.  Clearly something is missing here.
	
	Quasilocal energy is proportional to the extrinsic curvature.  However, for an arbitrary 2-manifold embedded in a 4-manifold, the extrinsic curvature has two components.  In other words, being of codimension two, $\Omega$ has two linearly independent normal vectors each of which has an extrinsic curvature associated with it.  A natural question then is to ask which of the two components should be used to calculate quasilocal energy.  Above, $\Omega$ is viewed as a submanifold of $\Sigma$, and its normal $n_i$ in $\Sigma$ is used to calculate the quasilocal energy.  However, one could have just as well viewed $\Omega$ as a submanifold of $B$ and used its normal $u_i$ in $B$ to perform the same calculation but with different results.  To put it differently, the calculation above depends in a critical way on the choice of foliation.  A different foliation of $M$ would mix the two components of extrinsic curvature, giving an entirely different expression for the quasilocal energy.  The static spacetimes considered in \cite{Brown:1992br} have a natural foliation for which the second component of extrinsic curvature vanishes and thus makes no contribution to the quasilocal energy.  For a Robertson-Walker spacetime, this is no longer the case.  What is missing from the above calculation is a proper treatment of both components of extrinsic curvature.

	\section{Quasilocal Energy \`a la Epp}
	\label{Quasilocal Energy a la Epp}
	The basic idea behind the notion of quasilocal energy, at least in the Brown-York and Epp formulations, is that the energy contained in a closed surface is related to the extrinsic curvature of that surface.  Therefore to begin with, a precise definition of extrinsic curvature is needed.
	
	Let $M$ denote an arbitrary manifold of dimension $m$, and $T_pM$ denote the tangent vector space of $M$ at a point $p$.  Let $T_pM$ be spanned by linearly independent vectors $\left\{ e_\mu \right\}$, where $\mu=1, \cdots, m$.  Let $g$ denote the metric and $\nabla$ the covariant derivative on $M$.
	
	Let $S$ denote a submanifold of $M$ of dimension $s$, and let $T_pS$ denote the tangent vector space of $S$ at the point $p$.  Let $T_pS$ be spanned by linearly independent vectors $\left\{ \xi_i \right\}$, where $i=1, \cdots, s$.  Let $N_pS$ denote the normal vector space of $S$ at the point $p$.  Let $N_pS$ be spanned by linearly independent vectors $\left\{ n_a \right\}$, where $a=1, \cdots, m-s$.
	
	An arbitrary vector $v \in T_pM$ can be decomposed into two components, one normal and the other tangent to $S$:
	\begin{align}
		v &= \left( v \right)^\bot + \left( v \right)^\| = \underbrace{\sum_{a=1}^{m-s} g \left( v,n_a \right) n_a}_{normal} + \underbrace{\left[ v - \sum_{a=1}^{m-s} g \left( v,n_a \right) n_a \right]}_{tangent}.
	\end{align}
	A similar thing can be done for the covariant derivative.  Given two arbitrary vectors $u,v \in T_pM$, the covariant derivative can be decomposed into normal and tangent components:
	\begin{align}
		\nabla_u v &= \left( \nabla_u v\right)^\bot + \left( \nabla_u v\right)^\| = \underbrace{\sum_{a=1}^{m-s} g \left( \nabla_u v,n_a \right) n_a}_{normal} + \underbrace{\left[ \nabla_u v - \sum_{a=1}^{m-s} g \left( \nabla_u v,n_a \right) n_a \right]}_{tangent}.
	\end{align}
	This decomposition can be written as $\nabla_u v = K (u,v) + \bar{\nabla}_u v$ where $K$ is the second fundamental form and $\bar{\nabla}$ is the induced covariant derivative on $S$.  The second fundamental form is in fact defined by this decomposition \cite{Bruhat:1982}:
	\begin{align}
		\label{second fundamental form}
		K \left( u,v \right) \equiv \sum_{a=1}^{m-s} g \left( \nabla_u v,n_a \right) n_a.
	\end{align}
	The induced covariant derivative is the projection of $\nabla_u v$ onto $S$, and is the unique covariant derivative that is torsion-free and compatible with the induced metric on $S$ \cite{Bruhat:1982}.
	
	The mean extrinsic curvature vector $H$ is defined as the trace of the second fundamental form:
	\begin{align}
		H \equiv \sum_{i=1}^{s} K \left( \xi_i,\xi_i \right),
	\end{align}
	where the vectors $\xi_i \in T_pS$ are tangent to the submanifold $S$.  The mean extrinsic curvature scalar $|H|$ is defined as the magnitude:
	\begin{align}
		\left| H \right| \equiv \sqrt{H \cdot H} \,\,.
	\end{align}
	To connect with the standard physics literature, it might be helpful to express some of these results in `indexed' notation.  The second fundamental form, mean extrinsic curvature vector, and mean extrinsic curvature scalar can be expressed respectively as:
	\begin{align}
		\label{second fundamental form indexed}
		K^\lambda_{\mu \nu} &= -\sum_{a=1}^{m-s} \nabla_\mu n_{a \nu} \,\, n^\mu_a \,\,, \\
		H^\lambda &= \sum_{i=1}^{s} K^\lambda_{\mu \nu} \, \xi^\mu_i \, \xi^\nu_i \,\,, \\
		\left| H \right| &= \sqrt{ g_{\mu \nu} \, H^\mu \, H^\nu } \,\,.
	\end{align}
	A minus sign appears in \eqref{second fundamental form indexed} when the derivative operator is moved to a neighboring term, and a projection operator should be used as appropriate when restricting these quantities to the submanifold $S$. 
	
	Note that according to its definition, the second fundamental form is a tensor of type $(1,2)$.  In the special case where $S$ is of codimension one, the second fundamental form is defined simply as $K \left( u,v \right) \equiv g \left( \nabla_u v,n \right)$.  In this case, since there is only one normal vector, it can be omitted from the definition, making the second fundamental form a tensor of type $(0,2)$.  In this context, the second fundamental form is also called the extrinsic curvature.  It can be shown \cite{Bruhat:1982} that $K = \left( -\frac{1}{2} \mathfrak{L}_n g \right)^\|$, where $\mathfrak{L}$ denotes the Lie derivative.  It can also be shown for $u,v \in T_pS$ and $\nabla$ compatible with $g$ that $K (u,v) = -g \left( v,\nabla_u n \right)$.  This is the basis for writing, as one often sees in the literature, $K_{ij} = - \nabla_i n_j$.  Also in this context, the mean extrinsic curvature vector is really just a scalar and reduces to the usual definition of mean extrinsic curvature found in the study of surfaces embedded in $\mathbb{R}^3$ for example.
	
	Adapting the above notation to the case at hand, $M$ becomes the four-dimensional spacetime, and $S$ becomes the two-dimensional boundary surface $\Omega$.  Being of dimension two and codimension two, $\Omega$ has two tangent vectors $\left\{ \xi_1^\mu,\xi_2^\mu \right\}$, and two normal vectors $\left\{ n_1^\mu,n_2^\mu \right\}$.  The Epp definition of quasilocal energy can be expressed in terms of the mean extrinsic curvature scalar.  The energy inside $\Omega$, without the reference term, is:
	\begin{align}
		\label{Epp QLE Definition}
		E_{phys} = -\frac{1}{8 \pi G} \int_\Omega d^2x \sqrt{|\sigma|} \,\, \left| H \right|.
	\end{align}
	A reference point must also be defined.  As before, Minkowski spacetime is used as the reference, and the Epp quasilocal energy becomes:
	\begin{align}
		\label{Epp QLE Definition 2}
		E_{E} &= E_{phys} - E_{ref} = -\frac{1}{8 \pi G} \int_\Omega d^2x \sqrt{|\sigma|} \,\, \left| H \right| + \frac{1}{8 \pi G} \int_{\bar{\Omega}} d^2\bar{x} \sqrt{|\bar{\sigma}|} \,\, \left| \bar{H} \right|.
	\end{align}
	Epp's derivation of this result was motivated by an analogy with the formula $E^2 = m^2 + p^2$ from special relativity, and so proceeded along different lines than that presented here.  However as Epp notes \cite{Epp:2000zr}, the final result can be expressed in terms of the means extrinsic curvature scalar exactly as in \eqref{Epp QLE Definition 2}.
	
	It is instructive to compare the Epp energy \eqref{Epp QLE Definition 2} to the Brown-York energy \eqref{BY QLE Definition 3}.  The main distinction of the Epp energy is that it is independent of the foliation.  This is so because in $|H|$ both components of the extrinsic curvature are accounted for and combined in a way that is invariant under a change of basis for the normal space $N_pS$.  Since the choice of foliation is tantamount to the choice of observers, this amounts to saying that the Epp energy is invariant under boosts of observers.  For this reason, it is tempting to refer to Epp's definition as quasilocal mass rather than energy.  However, other properties of this quantity compel Epp to refer to it as an `invariant quasilocal energy' rather than mass \cite{Epp:2000zr}.
	
	For stationary spacetimes, the Brown-York energy calculated for stationary observers is identical to the Epp energy.  To put it differently, suppose that for a stationary spacetime one chooses the natural constant-$t$ foliation, where $t$ is the coordinate along the timelike Killing vector.  Then, the Brown-York energy calculated for this foliation is equal to the Epp energy.  For example, both formulations results in the same expression for the quasilocal energy of Schwarzschild spacetime:
	\begin{align}
		\label{QLE schwarzschild}
		E_{BY} &= E_E = \frac{r}{G} \left( 1 - \sqrt{1 - \frac{2 G M}{r}} \right) \,\,.
	\end{align}
	In the limit $r \rightarrow \infty$, this result reduces to $M$ which is exactly the ADM energy.
	
	The real distinction between these formulations manifests itself when considering non-stationary spacetimes where the natural choice of foliation can result in non-vanishing extrinsic curvature in both components.  In the next section, this is demonstrated explicitly in the context of FRW cosmology.
	
	It should be mentioned that the Epp quasilocal energy \eqref{Epp QLE Definition 2} is similar to the definition offered by Wang and Yau in \cite{Wang:2008}.  The Wang-Yau construction requires a foliation and the choice of an admissible time function $\tau: \Sigma \rightarrow \mathbb{R}$.  (See \cite{Wang:2008} for precise definitions.)  Although their approach is considerably more complicated, it can be shown that with the natural foliation and a constant time function, $\tau: \Sigma \rightarrow t_0$, their approach and that of Epp give the same result when applied to the FRW universe.

	\subsection{FRW Cosmology \`a la Epp}
	\label{FRW Cosmology a la Epp}
	In this subsection I again calculate the quasilocal energy of an FRW universe, but this time using the Epp definition \eqref{Epp QLE Definition 2}.  As in section \ref{FRW Cosmology a la Brown and York}, consider the Robertson-Walker metric and choose $\Omega$ to be a sphere of comoving radius $r$ at time $t$.  The normal space of $\Omega$ is spanned by two unit vectors which, in coordinates $\left( t,r,\theta,\phi \right)$, are:
	\begin{align}
		n_{1}^{\mu} = (-1,0,0,0) \qquad , \qquad n_{2}^{\mu} = (0,\frac{\sqrt{1-\kappa r^2}}{a(t)},0,0).
	\end{align}
	These vectors are used to calculate the second fundamental form.  The tangent space of $\Omega$ is spanned by the following two unit vectors:
	\begin{align}
		\xi_1^\mu = \left( 0 , 0 , \frac{1}{a(t) r} , 0 \right) \qquad , \qquad \xi_2^\mu = \left( 0 , 0 , 0 , \frac{1}{a(t) r \sin\theta} \right)
	\end{align}
	The mean curvature vector and its magnitude are:
	\begin{align}
		H^\mu &= \left( \frac{-2 \dot{a}(t)}{a(t)} \,\, , \,\, \frac{-2 \left( 1-\kappa r^2 \right)}{r \, a^2(t)} \,\, , \,\, 0 \,\, , \,\, 0 \right) \\
		|H|      &= \frac{2 \sqrt{ 1-\kappa r^2 - r^2 \,\, \dot{a}^2(t)}}{r \,\, a(t)}.
	\end{align}
	The first term in \eqref{Epp QLE Definition 2} is then calculated to be:
	\begin{align}
		\label{Epp QLE FRW physical term}
		E_{phys} &= -\frac{1}{8 \pi G} \int_\Omega d^2x \sqrt{|\sigma|} \,\, \left| H \right| = -\frac{1}{G} r \,\, a(t) \,\, \sqrt{ 1-\kappa r^2 - r^2 \,\, \dot{a}^2(t)}.
	\end{align}
	
	To calculate the reference term in \eqref{Epp QLE Definition 2}, the same procedure as in section \ref{FRW Cosmology a la Brown and York} can be used.  Specifically, Minkowski spacetime is again the reference spacetime, and $\bar{\Omega}$ is defined by the same isometric embedding \eqref{isometric embedding ansatz}.  The normal space of $\bar{\Omega}$ is spanned by two unit vectors which, in coordinates $\left( \bar{t}, \bar{r}, \bar{\theta}, \bar{\phi} \right)$, are:
	\begin{align}
		\bar{n}_{1}^{\mu} = (-1,0,0,0) \qquad , \qquad \bar{n}_{2}^{\mu} = (0,1,0,0).
	\end{align}
	The tangent space of $\bar{\Omega}$ is spanned by the following two unit vectors:
	\begin{align}
		\bar{\xi}_1^\mu \equiv \left( 0 , 0 , \frac{1}{\bar{r}} , 0 \right) \qquad , \qquad \bar{\xi}_2^\mu \equiv \left( 0 , 0 , 0 , \frac{1}{\bar{r} \sin\bar{\theta}} \right)
	\end{align}
	The mean curvature vector and its magnitude are:
	\begin{align}
		\bar{H}^\mu &= \left( 0 \,\, , \,\, \frac{-2}{\bar{r}} \,\, , \,\, 0 \,\, , \,\, 0 \right) \\
		|\bar{H}|      &= \frac{2}{\bar{r}}.
	\end{align}
	The second term in \eqref{Epp QLE Definition 2} is then calculated to be
	\begin{align}
		\label{Epp QLE FRW reference term}
		E_{ref} &= - \frac{1}{8 \pi G} \int_{\bar{\Omega}} d^2\bar{x} \sqrt{|\bar{\sigma}|} \,\, \left| \bar{H} \right| = -\frac{1}{G} \bar{r} = -\frac{1}{G} r a(t).
	\end{align}
	Finally, subtracting \eqref{Epp QLE FRW reference term} from \eqref{Epp QLE FRW physical term}, the Epp quasilocal energy of a Robertson-Walker spacetime is found to be:
	\begin{align}
		\label{Epp QLE FRW}
		E_{E} &= E_{phys} - E_{ref} = \frac{1}{G} r \,\, a(t) \,\, \left[ 1 - \sqrt{ 1-\kappa r^2 - r^2 \,\, \dot{a}^2(t)} \right].
	\end{align}
	
	This new result, as opposed to \eqref{BY QLE FRW}, does not generally vanish for a flat, $\kappa = 0$, FRW model.  The quasilocal energy depends on both the scale factor and the rate of expansion.  These quantities are determined by the Friedmann equations and depend on details, such as matter density and equation of state, of the FRW model.
	
	Some obvious features of this result can be checked immediately.  Consider the case of a flat, $\kappa = 0$, static universe, $a(t)=1$.  In this case the FRW model reduces to the Minkowski spacetime, for which the quasilocal energy \eqref{Epp QLE FRW} is zero, as expected.  Of course, equations \eqref{Epp QLE FRW} and \eqref{BY QLE FRW} are equivalent for static universes.  The difference becomes apparent when the scale factor is allowed to evolve.  So consider the case where $\kappa = -1$ and $a(t) \sim t$, which describes the Milne universe.  It is well known that the Milne universe is just a portion of Minkowski spacetime expressed in different coordinates \cite{Carroll:2004st}.  Using \eqref{Epp QLE FRW} the quasilocal energy of the Milne universe is zero, as expected for any portion of Minkowski spacetime.  Equation \eqref{BY QLE FRW}, on the other hand, would not have given the correct result.

	\subsection{Comparison to Newtonian Cosmology}
	\label{Comparison to Newtonian Cosmology}
	A further test of the reliability of the result \eqref{Epp QLE FRW} is to compare it with a classical calculation of energy in Newtonian cosmology.  It is expected that in the case of a flat, matter-dominated universe, and in the small-sphere limit, the quasilocal energy \eqref{Epp QLE FRW} should reduce to the Newtonian results.
	
	For a flat universe ($\kappa = 0$) dominated by (non-relativistic) matter, the Friedmann equations \cite{Mukhanov:2005sc} imply that $\dot{a}(t) = \sqrt{ 2 A a^{-1}(t)}$, where $A \equiv \frac{4 \pi G}{3} \rho_{_{0}} a_{_{0}}^3$.  Here $\rho_{_{0}}$ and $a_{_{0}}$ are the matter density and scale factor at some initial time.  Using this, along with $\kappa =0$, in equation \eqref{Epp QLE FRW} results in:
	\begin{align}
		\label{Epp QLE flat matter}
		E_{E} &= \frac{1}{G} r \,\, a(t) \left[ 1 - \sqrt{ 1 - 2 A r^2 \,\, a^{-1}(t)} \right].
	\end{align}
	It is expected that in the small-sphere limit, i.e. as $r \rightarrow 0$, relativistic cosmology should resemble a Newtonian universe.  To examine this limit, \eqref{Epp QLE flat matter} should be expanded about $r=0$:
	\begin{align}
		\label{Epp QLE flat matter small sphere}
		E_{E} &= \frac{1}{G} a(t) \left[ A a^{-1}(t) r^3 + \frac{1}{2} A^2 a^{-2}(t) r^5 + O(r^7) \right] \nonumber\\
		       &= \frac{4 \pi}{3} \rho_0 a_0^3 r^3 + \frac{8 \pi^2 G}{9} \rho_0^2 a_0^6 a^{-1}(t) r^5 + O(r^7).
	\end{align}
	It is encouraging to see that the first term in this expansion corresponds to the rest mass energy of a spherical ball of gas.  To correctly compare this result with the Newtonian result, it should be noted that in FRW cosmology $r$ denotes a comoving coordinate, while in Newtonian cosmology $r$ usually denotes a fixed coordinate.  As a consequence, in FRW cosmology the proper distance from origin is $\sigma = a(t) r$, while in Newtonian cosmology the proper distance is simply $\sigma = r$.  Equation \eqref{Epp QLE flat matter small sphere}, expressed in terms of the proper distance, is:
	\begin{align}
		\label{Epp QLE flat matter small sphere proper}
		E_{E} &= \frac{4 \pi}{3} \rho_0 a_0^3 a^{-3}(t) \sigma^3 + \frac{8 \pi^2 G}{9} \rho_0^2 a_0^6 a^{-6}(t) \sigma^5 + O(\sigma^7).
	\end{align}
	
	Equation \eqref{Epp QLE flat matter small sphere proper} can now be compared to the Newtonian result.  A brief description of Newtonian cosmology can be found in \cite{Liddle:1998ew, Mukhanov:2005sc}, while more thorough accounts can be found in \cite{Bondi:1960, McCREA01011934, MILNE01011934}.  The appendix in this paper provides the necessary background for the purpose here.  Equation \eqref{total newtonian energy 2} from the appendix gives the total classical energy, including kinetic, potential and rest mass energy, inside a sphere of radius $r$:
	\begin{align}
		\label{total newtonian energy}
		E_{clas} &= \frac{4 \pi}{3} \rho_{_{0}} a_{_{N0}}^3 a_{_{N}}^{-3}(t) r^3 - \frac{2 \pi}{5} \rho_{_{0}} a_{_{N0}}^3 a_{_{N}}^{-5}(t) r^5 k,
	\end{align}
	where $k$ is a constant proportional to the mechanical energy, and $a_{_{N}}(t)$ is the Newtonian scale factor.\footnote{I distinguish between the FRW scale factor, $a(t)$, and the Newtonian one, $a_{_{N}}(t)$, because these two do not necessarily coincide.}  In terms of the proper distance, this becomes:
	\begin{align}
		\label{total newtonian energy proper}
		E_{clas} &= \frac{4 \pi}{3} \rho_{_{0}} a_{_{N0}}^3 a_{_{N}}^{-3}(t) \sigma^3 - \frac{2 \pi}{5} \rho_{_{0}} a_{_{N0}}^3 a_{_{N}}^{-5}(t) \sigma^5 k.
	\end{align}
	
	The difference between the quasilocal energy \eqref{Epp QLE flat matter small sphere proper} and classical energy \eqref{total newtonian energy proper} is then:
	\begin{eqnarray}
		\label{difference in energy}
		\Delta E & \equiv & E_{E} - E_{clas} \nonumber \\
		         &   =    & \frac{4 \pi}{3} \rho_0 \left[ a_0^3 a^{-3}(t) - a_{_{N0}}^3 a_{_{N}}^{-3}(t) \right] \sigma^3 \nonumber \\
		         &        & {} + \left[ \frac{8 \pi^2 G}{9} \rho_0^2 a_0^6 a^{-6}(t) + \frac{2 \pi}{5} \rho_{_{0}} a_{_{N0}}^3 a_{_{N}}^{-5}(t) k \right] \sigma^5 + O(\sigma^7).
	\end{eqnarray}
	The constant $k$ in Newtonian cosmology plays a role akin to $\kappa$ in FRW cosmology.  The case $k=0$ corresponds to an expanding gas in which every particle travels at exactly its escape velocity, allowing the gas to expand asymptotically to infinity.  To compare with a flat FRW universe I choose $k=0$ for which the Newtonian scale factor $a_{_{N}}(t)$ and the FRW scale factor $a(t)$ coincide exactly.  In this case, the difference \eqref{difference in energy} reduces to:
	\begin{eqnarray}
		\label{difference in energy 2}
		\Delta E & = & \frac{8 \pi^2 G}{9} \rho_0^2 a_0^6 a^{-6}(t) \sigma^5 + O(\sigma^7).
	\end{eqnarray}
	Thus this calculation of quasilocal energy suggests that there are corrections to the classical energy appearing at order $O(\sigma^5)$.  One can be even more explicit by substituting for the scale factor which, for a flat matter-dominated universe, is $a(t) \sim t^{2/3}$.  The correction \eqref{difference in energy 2} then becomes $\Delta E \sim t^{-4} \sigma^5 + O(\sigma^7)$.  This then suggests that although there are corrections to the classical energy, they tend to vanish quickly with time.

	\section{Conclusion}
	Defining a universal notion of energy has been a challenge since the inception of general relativity.  Today a myriad of definitions exist.  Although it is generally agreed that energy ought to be a quasilocal quantity, it is not at all clear which of the many quasilocal definitions is the correct one.  Some even question whether a correct definition exists.  Perhaps, like so many important questions in physics, this one ought to be settled by experiment or observation.
	
	The field of cosmology has proven to be a fertile testing ground for many of the ideas in general relativity.  So it seems appropriate to look to cosmology for clues to the nature of energy.  The primary goal of this paper has been to calculate the quasilocal energy of an FRW universe, the so-called standard model of cosmology.  For this purpose two specific definitions of quasilocal energy were chosen.  It has been shown that in FRW cosmology, or more generally for non-stationary spacetimes, a clear distinction arises between the Brown-York and Epp formulations.  It seems that Epp's invariant quasilocal energy (or mass) is more appropriate for describing what we normally think of as cosmological energy.  For non-stationary spacetimes, there is little precedence, and therefore the validity of equation \eqref{Epp QLE FRW} can be examined only in special limits.  The comparison with Newtonian cosmology shows that the results here are reliable at least to lowest order.  The presence of higher order corrections to Newtonian cosmology is of course expected.  It would be interesting to examine how these corrections manifest themselves in cosmological data.  It is conceivable that the presence of this additional energy may affect structure formation in the early universe, alter the rate of expansion, or manifest itself as dark energy \cite{Wiltshire:2007zh}.
	
	This work can be extended in several directions.  The most immediate extension would be to consider other formulations of quasilocal energy.  The constructions by Hawking \cite{Hawking:1968qt}, Kijowski \cite{Kijowski:1997}, and by Liu and Yau \cite{Liu:2003bx} are similar in structure to the two cases studies here.  Several special cases of the FRW model, such as de Sitter (dS) and anti-de Sitter (AdS) spacetimes, deserve closer examination.  In fact, using appropriate coordinates, the metrics for these spacetimes can be written in the form of the Robertson-Walker metric.  So the result \eqref{Epp QLE FRW} is already valid for dS and AdS so long as the coordinates are carefully considered.  The thermodynamics of these spacetimes is dependent on energy, and thus provides another interesting application of quasilocal energy.

	\section*{Acknowledgements}
	I would like to thank Steven Carlip for offering advice and guidance without which this paper would not have been possible.  This work was supported in part by the Department of Energy under grant DE-FG02-91ER40674.

	\section*{Appendix:  Newtonian Cosmology}
	Here I provide a review of Newtonian cosmology sufficient for deriving equation \eqref{total newtonian energy}.
	
	The setting for Newtonian cosmology is flat, 3-dimensional, euclidean space.  I will use spherical coordinates denoted by $\mathbf{r}=\left( r,\theta,\phi \right)$.\footnote{Here $\mathbf{r}$ is a fixed coordinate, not to be confused with the comoving coordinate used in FRW cosmology.}  First, it is assumed that space is filled with dust of homogeneous density $\rho(t)$ out to some large radius $R(t)$.  Note that both the density and the radius depend on time, indicating that the ball of dust may be expanding or contracting.  Second, conservation of mass is assumed, and in particular that the continuity equation holds:  $\nabla \cdot \mathbf{J}=-\partial \rho / \partial t$.  Third, it is assumed that, inside the radius $R(t)$, Hubble's law holds:  $\mathbf{v} \left( \mathbf{r},t \right)=H(t) \mathbf{r}$, where $H(t)$ is the Hubble parameter.
	
	Hubble's law can be expressed as $\dot{\mathbf{r}}=H \mathbf{r}$.  This equation can be solved to obtain:
	\begin{equation}
		\label{rH}
		r(t) = r_0 \exp \left[ \int_{t_0}^{t} H \left( t' \right) dt' \right],
	\end{equation}
	where $r_0 \equiv r \left( t_0 \right)$ for some initial time $t_0$.  Using the definition $\mathbf{J} \equiv \rho \mathbf{v}$ and the homogeneity of density, the continuity equation can be expressed as $\rho \nabla \cdot \mathbf{v} = -\dot{\rho}$.  Using Hubble's law in this last equation, it can be modified to $3 \rho H = -\dot{\rho}$, which then can be solved to obtain:
	\begin{equation}
		\label{rhoH}
		\rho (t) = \rho_0 \exp \left[ -3 \int_{t_0}^{t} H \left( t' \right) dt' \right],
	\end{equation}
	where $\rho_0 \equiv \rho \left( t_0 \right)$.  The Newtonian scale factor $a_{_{N}}(t)$ is defined by $H(t) = \dot{a}_{_{N}}(t)/a_{_{N}}(t)$.\footnote{The scale factor in Newtonian cosmology is not necessarily the same as in FRW cosmology, and is thus distinguish here by the subscript N.}  This equation can be solved to obtain:
	\begin{equation}
		\label{aH}
		a_{_{N}}(t) = a_{_{N0}} \exp \left[ \int_{t_0}^{t} H \left( t' \right) dt' \right],
	\end{equation}
	where $a_{_{N0}} \equiv a_{_{N}} \left( t_0 \right)$.  This last equation can be used to replace $H(t)$ with $a_{_{N}}(t)$ in equations \eqref{rH} and \eqref{rhoH}:
	\begin{eqnarray}
		\label{ra} r(t) & = & r_0 \frac{a_{_{N}}(t)}{a_{_{N0}}} \\
		\label{rhoa} \rho(t) & = & \rho_0 \left[ \frac{a_{_{N}}(t)}{a_{_{N0}}} \right]^{-3}
	\end{eqnarray}
	
	Consider a test mass $m$ located at a distance $r < R$ from the origin.  The force on this test mass, according to Newtonian mechanics, is the gravitational force of the mass enclosed inside the radius $r$; the mass outside that radius has no effect on the test mass.  So the force on test mass $m$ is:
	\begin{align}
		\label{newtonian force}
		F = \frac{GMm}{r^2},
	\end{align}
	where $M = \frac{4}{3} \pi r^3 \rho(t)$ is the total mass inside radius $r$.  Writing this equations as $\ddot{r} = -\frac{4 \pi G}{3} r \rho$, and then using equations \eqref{ra} and \eqref{rhoa} to eliminate $r(t)$ and $\rho(t)$ in favor of $a_{_{N}}(t)$, results in:
	\begin{equation}
		\label{adoubledot}
		\ddot{a}_{_{N}}(t) = -\frac{4 \pi G}{3} \rho_0 a_{_{N0}}^3 a_{_{N}}^{-2}(t).
	\end{equation}
	Equation \eqref{adoubledot} can be integrated to obtain
	\begin{equation}
		\label{asingledot}
		\dot{a}_{_{N}}(t) = \sqrt{ \frac{8 \pi G}{3} \rho_{_{0}} a_{_{N0}}^3 a_{_{N}}^{-1}(t) - k },
	\end{equation}
	where $k \equiv \frac{8 \pi G}{3} \rho_{_{0}} a_{_{N0}}^{2} - \dot{a}_{_{N0}}^2$ is a constant.  This equation can be rearranged and written in a form that resembles one of Friedmann's equations:
	\begin{align}
		\left( \frac{\dot{a}_{_{N}}}{a_{_{N}}} \right)^2 = \frac{8 \pi G}{3} \rho - \frac{k}{a_{_{N}}^2}.
	\end{align}
	
	To discuss energy in the context of Newtonian cosmology, consider again the test mass $m$ located at distance $r < R$.  From equation \eqref{newtonian force}, the gravitational potential energy of mass $m$ is calculated to be:
	\begin{align}
		V_m = -\frac{GMm}{r} = -\frac{4 \pi G }{3} \rho r^2 m.
	\end{align}
	The kinetic energy of the same test mass is:
	\begin{align}
		T_m = \frac{1}{2} m v^2 = \frac{1}{2} m H^2 r^2.
	\end{align}
	Defining the mechanical energy of mass $m$ as $U_m \equiv T_m + V_m$ results in:
	\begin{align}
		\label{mechanical energy}
		U_m = \frac{1}{2} m H^2 r^2 - \frac{4 \pi G }{3} \rho r^2 m.
	\end{align}
	Assuming uniform density, the total mechanical energy inside a sphere of radius $r < R$ can be found by integrating \eqref{mechanical energy} over the volume of the sphere:
	\begin{align}
		\label{mechanical energy total}
		U &= \int \left( \frac{1}{2} \rho H^2 r^2 - \frac{4 \pi G }{3} \rho^2 r^2 \right) d^3x = \frac{2 \pi}{5} \rho H^2 r^5 - \frac{16 \pi^2 G}{15} \rho^2 r^5.
	\end{align}
	To find the \emph{total} energy, one should add to mechanical energy the rest mass energy as well.  So the total classical energy, including kinetic, potential and rest mass, is:
	\begin{align}
		\label{total newtonian energy 2}
		E_{clas} &= \frac{4 \pi}{3} \rho r^3 + \frac{2 \pi}{5} \rho H^2 r^5 - \frac{16 \pi^2 G}{15} \rho^2 r^5 \nonumber \\
		           &= \frac{4 \pi}{3} \rho_{_{0}} a_{_{N0}}^3 a_{_{N}}^{-3} r^3 + \frac{2 \pi}{5} \rho_{_{0}} a_{_{N0}}^3 \dot{a}_{_{N}}^2 a_{_{N}}^{-5} r^5 - \frac{16 \pi^2 G}{15} \rho_{_{0}}^2 a_{_{N0}}^6 a_{_{N}}^{-6} r^5 \nonumber \\
		           &= \frac{4 \pi}{3} \rho_{_{0}} a_{_{N0}}^3 a_{_{N}}^{-3} r^3 - \frac{2 \pi}{5} \rho_{_{0}} a_{_{N0}}^3 a_{_{N}}^{-5} r^5 k,
	\end{align}
	where in the second line I have used equation \eqref{rhoa}, and in the third line I have used the definition of $k$.
	
	It should be pointed out that $k$ is proportional to the mechanical energy of a unit mass:
	\begin{align}
		k = -2 \left( \frac{a_{_{0}}}{r_{_{0}}} \right)^2 U.
	\end{align}
	The mechanical energy is a conserved quantity in Newtonian cosmology, which explains why $k$ is referred to as a constant.  Most treatments of Newtonian cosmology \cite{Bondi:1960, MILNE01011934, Mukhanov:2005sc} consider only the case where $U=0$ and thus $k=0$.  This corresponds to an expanding gas in which every particle travels at exactly its escape velocity, allowing the gas to expand asymptotically to infinity.  The case $k<0$ corresponds to an expanding gas with speeds less than the escape velocity, resulting ultimately in contraction of the gas under the influence of gravity.  The case $k>0$ corresponds to an expanding gas with speeds greater than the escape velocity, resulting in eternal expansion.

\end{document}